\documentclass[preprint,review,times,3p]{elsarticle}
\usepackage{graphicx}
\usepackage{subfig}
\usepackage[countmax]{subfloat}
\usepackage{amsmath, amssymb}

\biboptions{sort&compress,comma}

\journal{ Physica E}

\begin{document}

\begin{frontmatter}
\title{Screening theory based modeling of the quantum Hall based quasi-particle interferometers defined at quantum-dots}
\author[label1]{A. Salman\corref{label5}}
\ead{aysevil@akdeniz.edu.tr} \cortext[label5]{Corresponding
Author.
Tel.: +902423102288; Fax: +902422278911} 
\author[label1]{E. Koymen}
\author[label1]{M. B. Yucel}
\author[label2]{H. Atci}
\author[label2]{U. Erkarslan}
\author[label3,label4]{A. Siddiki}
\address[label1]{Akdeniz University, Faculty of Sciences, Physics Department, Antalya 07058, Turkey}
\address[label2]{Mugla University, Faculty of Sciences, Physics Department, Kotekli 48170, Mugla, Turkey}
\address[label3]{Istanbul University, Faculty of Sciences, Physics Department, Vezneciler-Istanbul 34134, Turkey}
\address[label4]{Harvard University, Physics Department, Cambridge 02138 MA, USA}

\begin{abstract}
In this work, we investigate the spatial distributions and the widths of the
incompressible strips, \texttt{i.e.} the edge-states. The incompressible
strips that correspond to $\nu=1, 2$ and $1/3$ filling factors are examined in the
presence of a strong perpendicular magnetic field. We present a
microscopic picture of the fractional quantum Hall effect based
interferometers, within a phenomenological model. We adopt Laughlin quasi-particle
properties in our calculation scheme. In the
fractional regime, the partially occupied lowest Landau level is
assumed to form an energy gap due to strong correlations. Essentially by including this energy
gap to our energy spectrum, we obtain the properties of the
incompressible strips at $\nu=1/3$. The interference conditions
are investigated as a function of the gate voltage and steepness
of the confinement potential, together with the strength of the
applied magnetic field.
\end{abstract}

\end{frontmatter}

\section{Introduction}
\label{intro} The quantum behavior of particles has attracted many
concern; one of these concern is the quantum Hall based
interferometers which are realized in the two dimensional electron
systems (2DESs). The discovery of the
integer~\cite{vKlitzing80:494} and fractional~\cite{Tsui82:FQHE}
quantum Hall effects paved the path to investigate the importance
of interactions, in particular, by intensive efforts related with
the transport phenomena in the 2DES. At low temperatures and at certain
 magnetic field intervals, it is observed that the Hall
conductance $\sigma_{xy}$ assumes quantized values in units of
$e^2/h$, whereas the longitudinal conductance
$\sigma_{xx}$ vanishes. The experimental realization of
the integer quantum Hall effect (IQHE) shows that the
constant Hall conductance occurs if the electron number density
$n_{\mathrm{el}}$ is an integer ($i$) multiple of the magnetic
flux density $n_{\Phi_0}$. This ratio is named as the filling
factor, $\nu=n_{\mathrm{el}}h/eB$ and it describes the Landau level
filling at a given magnetic field. In the integer regime, i.e. if
$\nu=i$, it means that the Fermi energy falls between the two
adjacent Landau levels and the electrostatic potential cannot be
screened, the electron density remains constant. This state is
called incompressible. If Fermi energy equals to a Landau level,
the screening ability is good and this state is called
compressible. In the fractional quantum Hall effect (FQHE) regime, the
Landau level is partially occupied, hence, $\nu=f$. At certain fractional filling
factors, it has been shown that this partially occupied highly
degenerate Landau level splits into degenerate sub levels, and
energy gaps are formed. These gaps lead also an incompressible state.

In 1982, Wilczek introduced a new kind of statistical particles
named anyons~\cite{Wilczek82:1144,Wilczek82:957}, which is
described by fractional statistics in two dimensions. The quantum
mechanics of anyons are formulated by including a statistical
interaction potential to the Lagrangian or Hamiltonian of ordinary
particles~\cite{ezawa}. According to the quantum field theory of
anyons, the electromagnetic potential is replaced by the
Chern-Simons field. The Chern-Simons field is defined by the
composite-particle (fermion or boson) field that represents
anyons. In the case of a two dimensional electron gas subject to
strong perpendicular magnetic field, these anyonic particles are
 called Laughlin quasi-particles (LQPs) which obey fractional statistics and have
 an effective fractional electric charge $e^*=e/(2i+1)$. Laughlin
describes the Landau level filling for FQH states as $f=1/(2i+1)$
where $i$ is an integer~\cite{laughlin83} and LQPs are described by
elementary charged excitations of FQH
condensate~\cite{goldman07:e/3}.

The ``quantum Hall'' based interferometers, has revealed a novel
technique to exploit the properties of the quasi-particles, which
utilizes the so called edge states (ESs) in the extreme quantum
limit. The ESs are considered as monochromatic (-energetic) beams
that carry quasi-particles without scattering. Therefore, the
overall interference pattern also strongly depends on the spatial
distribution of these states. To understand the properties of
Laughlin quasi-particles, a number of experiments has been
performed by Camino et al.
~\cite{goldman07:e/3,Goldman05:155313,goldman06:Transport}. A key
element of these experiments are the quantum point contacts (QPCs)
and the electrostatic potential profile near these QPCs together
with the quantum dot. Electrostatics play an important role on
the formation and the spatial rearrangement of the electrostatic
potential. Moreover, the interaction of the electrons or
quasi-particles was proposed to be a possible origin of the
dephasing and a better understanding requires a self-consistent
(SC) calculation of the electrostatic potential ~\cite{siddiki07}.
Here we present an implementation of the SC-Thomas-Fermi-Poisson
approach to investigate the effects induced by direct Coulomb
interactions considering an homogeneous 2DES, where the interferometer
is defined by a quantum dot.

\section{The Model}\label{model}
The electrostatic treatment of the ESs is handled analytically by
Chklovskii, Shklovskii and Glazman (CSG)~\cite{Chklovskii92:4026}. In their pioneering work, they developed an electrostatic
model considering the gate defined samples and studied the formation of edge states. Using
the stability condition of the electrostatic potential, they
obtained the positions and the widths of the incompressible
states, which resemble the edge states. They argued that their
results can be employed to the etched structures by using
appropriate gate voltages that simulating etched edges ~\cite{Chklovskii92:4026}. Subsequently,
Gelfand and Halperin (GH) studied the edge electrostatics and
investigated the distribution of electron density considering an
etching defined sample ~\cite{Halperin94:etchedge}. Both of the
theories are accepted to be viable also for FQHE
~\cite{Chklovskii92:4026,Halperin94:etchedge}. In a recent work, the positions and the widths of incompressible states
are studied by Salman et al for both cases investigating a regular Hall bar, utilizing a self-consistent numerical approach ~\cite{salman:2010}. There, it is shown that the predictions of the charge density profile within the analytical approximations deviate considerably from the self-consistent calculations.

In this work, we employ the commonly used self-consistent
Thomas-Fermi-Poisson approach (SCTFPA) to obtain the electrostatic
potential and the electron distribution in the structures. The
total potential energy that electron experiences, within a
mean-field approximation, is \begin{equation}
V(x,y,z)=V_{d}(x,y,z)+V_{g}(x,y,z)+V_{\mathrm{surf}}(x,y,z)+V_{\mathrm{int}}(x,y,z),
\end{equation} here, $V_{d}(x,y,z)$ is generated by the ionized
donors, $V_{g}(x,y,z)$ is due to metallic gates deposited on the
surfaces, $V_{\mathrm{surf}}(x,y,z)$ represents the surface
potential and the last term stands for the electron-electron
interactions, which is determined by the electron density distribution $n_{\mathrm{el}}(x,y,z)$. The surface and gate potentials are fixed by the sample properties, e.g. the surface potential is determined by the mid-gap energy of the heterostructure. The potential generated by the charges are calculated via solving the Poisson's equation for given boundary conditions numerically.

In our self-consistent calculation scheme, we fix the homogeneous
immobile charge distribution, i.e. the donors; then the electron
density profile is calculated by the electrostatic potential
emanating from the donors, surface and gates. Next, the
electrostatic potential is determined by the electron density
profile in a self-consistent loop, at zero magnetic field and temperature. To start the self-consistent calculation we focus on the
lithographically defined sample, resembling the experimental structures
~\cite{goldman07:e/3,Goldman05:155313,goldman06:Transport}. By
considering the crystal growth parameters together with the
surface image of the quantum dot pattern, we calculate the charge
distribution at the 2DES. The details of the self-consistent
process are given in Ref.~\cite{siddiki04:195335,siddiki03:125315,Sefa08:prb}.


\begin{figure}[h!]
\begin{center}\leavevmode
\includegraphics[width=13cm,height=9cm]{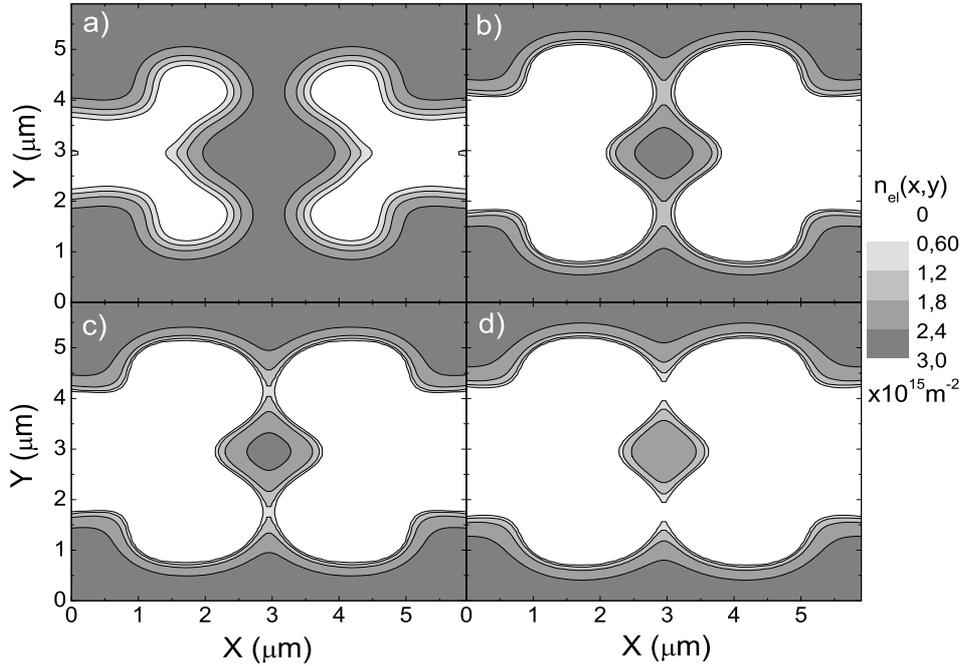}
\vspace{-0.2cm} \caption{ (Color online) Spatial distribution of
electrons at zero magnetic field and zero temperature. (a) gated
sample where potential is fixed to $V_g=-2.0V$.
 (b) and (c) etched structures, whose etching depths are 160 nm and
 240 nm from the surface of the sample, respectively. (d) etched
 and also gated (trench gated) sample: etching depth is 240 nm,
  gate voltage is $-2.0V$.\label{fig:figure1}}
\end{center}
\end{figure}

 The next step
is to determine the spatial positions of the edge states, which is
a straightforward procedure when the electron density
distribution is known. Our self-consistent scheme provides this
information, and integer or fractional filling factors (the edge
states) can be found simply by
$\nu(x,y)=2\pi\ell^2n_{\mathrm{el}}(x,y,z=\mathrm{2DES})$; here
 $\ell = \sqrt{\hbar/eB}$ is the magnetic length. In a further step, one can calculate also the
widths of the incompressible edge states, by following the
pioneering work of Chklovskii et al ~\cite{Chklovskii92:4026} given as
\begin{equation} a_{i,f}=\sqrt{\frac{2\epsilon \Delta E_{i,f}
}{\pi^2 e^2 {{dn}/{dx}}|_{x=x_{i,f}}}} \end{equation} where
$\Delta E_{i,f}$ is the energy gap, assuming the Zeeman split
Landau gap for the integer case and the many-body effects
induced gap for the fractional case.


\begin{figure}[h!]
\begin{center}\leavevmode
\includegraphics[width=13cm,height=9cm]{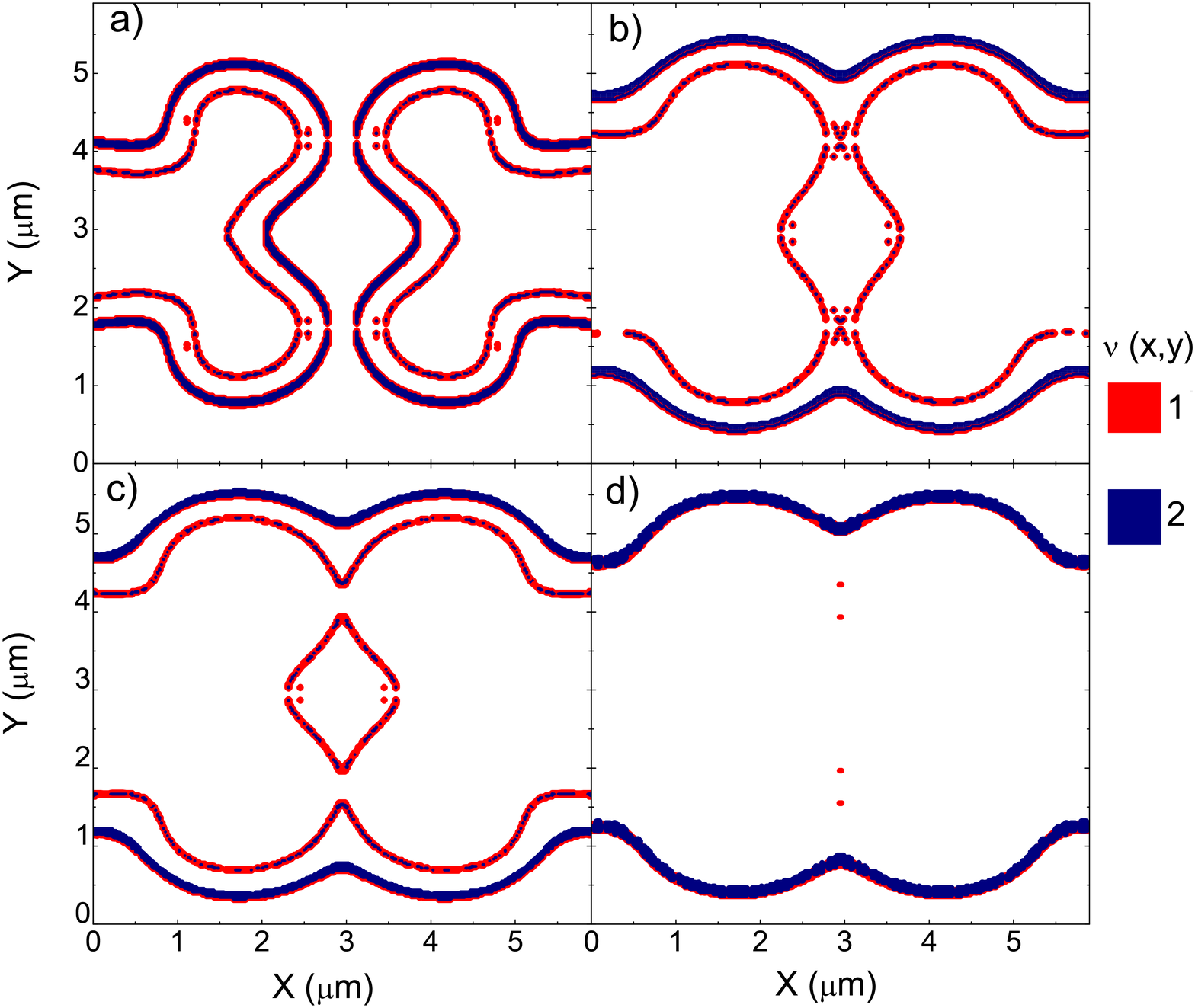}
\vspace{-0.2cm} \caption{ (a) and (b) show the spatial distribution
of ISs for gated samples at 2.6 T magnetic field. $V_g=-2.0V$ in (a)
and $V_g=-5.0V$ in (b). (c) and (d) show the spatial distribution of
ISs for 160nm etched and also -2.0V potential is applied (trench
gated) sample. The magnetic field is 2.6 T in (c) and 5.1 T in
(d).\label{fig:figure2}}
\end{center}
\end{figure}

Aharonov-Bohm interferometers (ABI) are basically used to measure
the vector potential induced phase difference between two paths of the (quasi-)particles.
The single electron states at the Landau level are quantized by the
Aharonov-Bohm condition and determines the phase, whereas for the LQPs the phase is determined by the properties of the composite particle~\cite{Goldman05:155313}. In the existence
of a perpendicular magnetic field, the closed paths define an area
$S=\pi r^2$ with radius $r$ that encircles the magnetic flux
$\Phi$ and is given by $\Phi = BSm = m\Phi_0$. The encircled area of the orbital is given by $S_m$ and $m$ represents the quantum numbers of the
orbital, where $\Phi_0=h/e$ is the magnetic flux
quantum~\cite{Goldman05:155313}. The theoretical predictions by
Aharonov and Bohm~\cite{Aharonov:Bohm} states that the interference
between the two paths depends on the phase difference, and the phase
difference is given by the integer multiples of the magnetic flux
quanta $\Phi$ encircled. Therefore, the area of an orbit is $S_m =
2\pi m\ell^2$ where $\ell$ is the radius of the
orbit~\cite{cicek10}. The region between the two adjacent orbits is $S_{m+1} - S_m = h/eB$, and this is related with the occupation of these orbits which is given by the filling factor as described above,
$i. e.$  $\nu=2\pi \ell^2n_{\mathrm{el}}$. Therefore, it is
essential to find the exact locations of integer or fractional
edge states to pin down the area enclosed by the interference paths. In the following section, the spatial distributions of the edge states will be provided by considering real experimental parameters.

\section{Results and Discussion}

The physical area of the structure that we considered is 5.9
$\mu$m$\times$ 5.9 $\mu$m, and the 2DES is 280 nm below the
surface in $z$ direction, similar to the experimental case
~\cite{goldman07:e/3,Goldman05:155313,goldman06:Transport}. The
2DES is generated at the interface of the GaAs-AlGaAs
heterostructure. Hence, we set the Lande-g* factor to be -0.44, as
default. We obtain the electron densities self-consistently as
mentioned above. Typical electron density distributions are shown
in Fig~\ref{fig:figure1}, (a) considering gate, (b) shallow etched, (c) deep etched and (d) trench gate (i.e. first etched then gated) defined samples. The gray scale depicts the electron occupied regions, whereas white areas denote the electron depleted regions. Here, the physical dimensions of the surface pattern are taken to be the same.

Since the electron free (white) regions are dominant in (b-d) one can conclude from Fig.~\ref{fig:figure1} that, etching is more
effective in depleting electrons when it is compared to solely gate defined quantum dots. Note that, to manipulate the electron density distribution, hence the area, one needs to impose metallic gates biased by negative voltages. Therefore, we also show the trench gated samples, resembling the experiments. This method is the most powerful technique to investigate both the edge effects and interference phenomena. In our calculations, we follow the arguments of Gerhardts~\cite{Guven03:115327}, Fogler~\cite{Fogler94:1656} and Chang~\cite{chang} in describing the non-equilibrium current. We assume that the current is carried by incompressible regions, since scattering is completely suppressed due to the lack of available states at the Fermi energy. However, our calculations are also viable if one utilizes the Landauer-B\"uttiker picture for transport and considers solely the capacitive effects proposed by Halperin and his co-workers~\cite{buttiker86,neder}.


\begin{figure}[h!]
\begin{center}\leavevmode
\includegraphics[width=13cm,height=9cm]{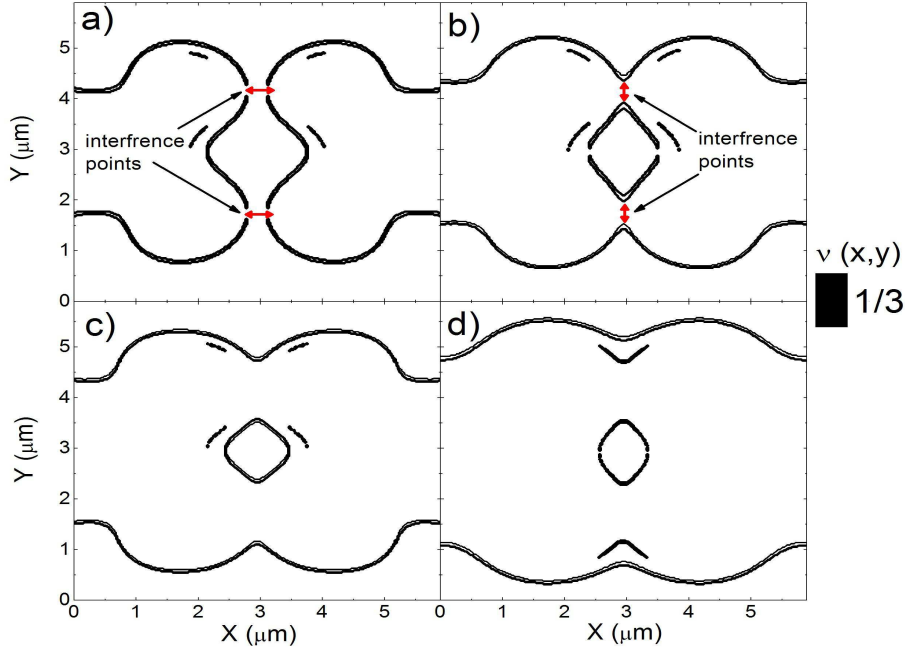}
\vspace{-0.2cm} \caption{Spatial distribution of the
incompressible strips for $\nu=1/3$ state at different gate
voltages. In (a) $V_g=-3.0V$, (b) $V_g=-4.0V$, (c) 160 nm etched
sample is considered at $14.4$ T (d) 160 nm etched and also
$V_g=-4.0V$ gated sample is considered at $14.4$ T.
\label{fig:figure3}}
\end{center}
\end{figure}

Fig.~\ref{fig:figure2} shows the ESs considering the filling
factors $\nu=1$ and $2$. To find the
spatial distribution of the incompressible strips, we imposed magnetic field values 2.6 T and 5.1 T, that are consistent with the
experiments~\cite{Goldman05:155313}. As it is seen from
Fig.~\ref{fig:figure2}, there are two edge channels (inner (light)
and outer (dark)). The outer edge channel corresponds to $\nu=2$
and inner edge channel corresponds to $\nu=1$. At relatively low
gate voltages, Fig.~\ref{fig:figure2}a, the both edge channels are
transmitted through the constrictions, whereas at higher negative
gate voltages (Fig.~\ref{fig:figure2}b) the outer edge state is
back-reflected; meanwhile, the inner edge states come closer to
each other and start to overlap. Thus, the interference is more
possible for lower gate voltages, at least for the inner channel.
In Fig.~\ref{fig:figure2}b, we observe that the inner channels
overlap, whereas the outer channels are (back-)reflected, hence, at this
configuration tunneling becomes impossible for $\nu=2$ state.
Fig.~\ref{fig:figure2}c and Fig.~\ref{fig:figure2}d depict the results for trench gated samples (160 nm etched and -2.0 V is applied to the gates). The outer edge channels are decoupled, hence interference is suppressed.
Also in Fig.~\ref{fig:figure2}d, the inner edge channels disappear
due to the steepness of the electron density that is considered for trench
gated samples. The incompressible strip is very narrow compered to the
magnetic length under these conditions, and the inner edge channel is therefore
suppressed. For high magnetic fields none of the channels are
transmitted, therefore, interference is completely washed out.

We also investigate the $\nu=1/3$ edge state in the presence of
strong magnetic fields. We showed that it is more possible to
observe the interference at $\nu=1/3$, particularly at gated samples by
imposing low gate voltages, Fig.~\ref{fig:figure3}a. When the gate
voltages are increased, electrons are repelled more from the
edges, therefore, the edge states on the left and right start to
overlap. Etching also depletes the electrons more sharply than
gating as a result of steeper potential gradient, hence, the
quasi-particle edge states overlap. If the sample is made by
the trench gated as shown in Fig.~\ref{fig:figure3}d, it is observed
that the edge states are dispelled the most effectively.

\begin{figure}[h!]
\begin{center}\leavevmode
\includegraphics[width=13cm,height=9cm]{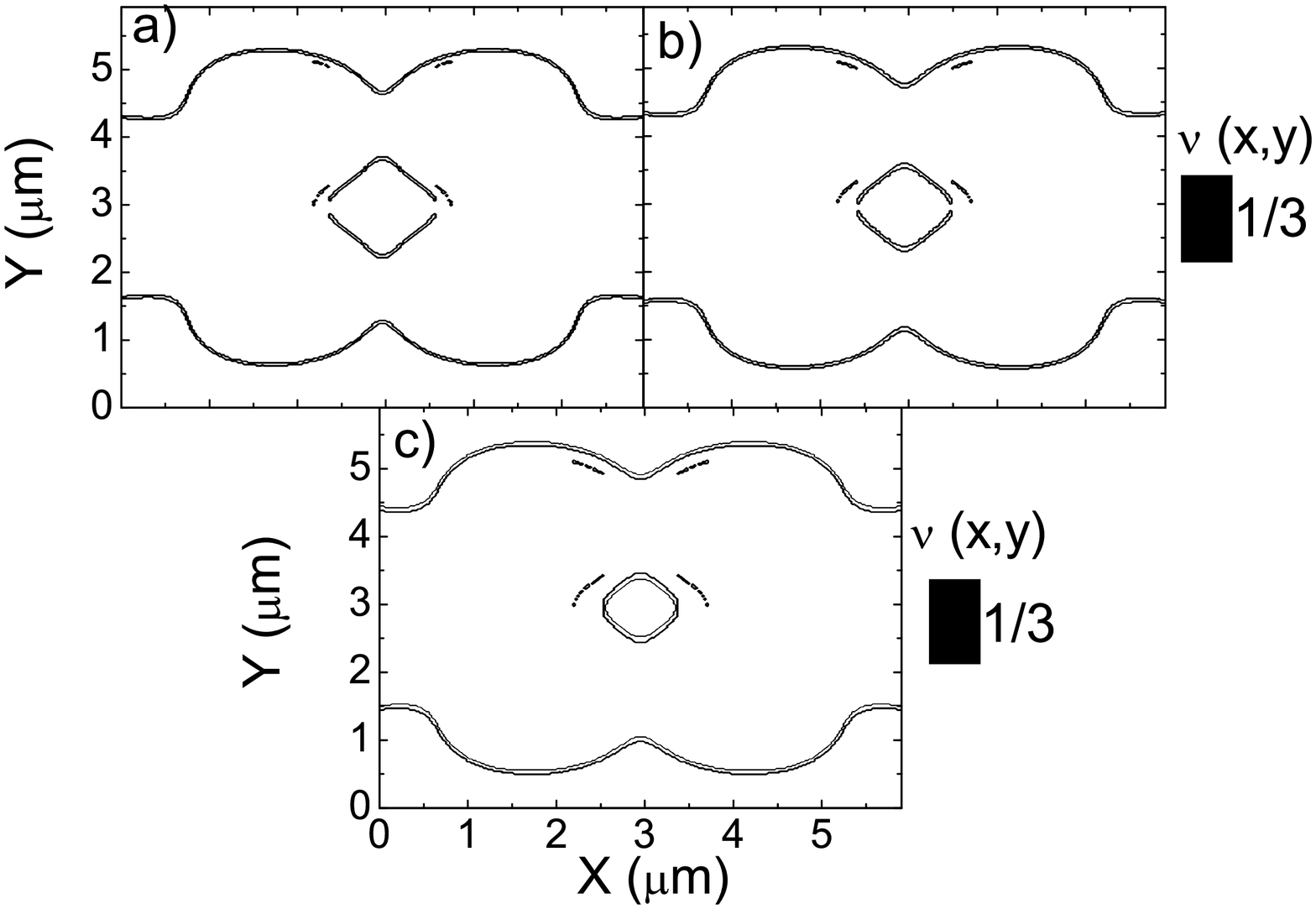}
\vspace{-0.5cm} \caption{Spatial distribution of the incompressible
strips for $\nu=1/3$ state in 240 nm etched sample with different
magnetic field
 values: (a) B=12.0 T, (b) B=13.2 T (c) B=14.4 T. \label{fig:figure4}}
\end{center}
\end{figure}

In a further investigation, we studied the effect of magnetic
field on the path of the edge states, hence the enclosed area. The
distribution of $\nu=1/3$ edge state is shown in
Fig.~\ref{fig:figure4}. The calculated areas of the enclosing
loops in Fig.~\ref{fig:figure4} are $1.04~\mu$m$^2$ for (a),
$0.78~\mu$m$^2$ for (b) and $0.45~\mu$m$^2$ for (c). It is
interesting to observe that, by changing magnetic field monotonously, the actual area enclosed by the $\nu=1/3$ edge state changes approximately by a factor of 2. Therefore, assuming an area independent charging model seems not to be plausible. However, one should also keep in mind that in actual experiments, the interference signal is not observed in a large $B$ interval. This is due to the fact that not only the area is changed, but also the tunneling mechanisms are altered. To be explicit, while the magnetic field varied, one also suppresses the scattering between the edge states by pulling them apart. Hence, no partitioning can take place and interference signal disappears. Therefore, one cannot measure the phase difference between two paths. This observation, of course, does not hinder the fact that the area enclosed is strongly affected by the change of magnetic field.

\section{Conclusions}
\label{conc} In this work, we numerically investigated the
electrostatics of a field effect induced quantum dot, and hence, the spatial distribution of the edge
states by considering the Aharonov-Bohm interference experiments
performed in the integer ($\nu=1, 2$) and fractional ($\nu=1/3$)
quantized Hall regimes~\cite{Goldman05:155313,goldman07:e/3}. We
observed that, defining the interference device (i.e. the quantum
dot) which is created by trench gating provides a steeper edge potential profile, and therefore, this device is more effective for interferometric measurements. In contrast, the gate defined devices enables higher visibility at
fractional states, due to the fact that these states become more
stable since the density gradients are smoother. Another
observation is related with the change of the area while sweeping
the magnetic field. At the integer regime, edge states
are less affected by the change in $B$, hence capacitive models
that essentially neglect the areal dependency may become more reliable. Our final remark for the fractional case is, the area changes approximately a factor of 2 while changing the magnetic field about 2.5 Tesla. Hence, an areal independent model is more questionable.

The observations yield the following conclusion: To investigate
the importance of the areal dependency, therefore the importance
of Hartree type interactions, it is necessary to perform experiments with the same sample geometry for the both gate and trench gated samples. We expect to see that the gate defined samples will show an enhanced areal dependency, due to the fact that their edge potential profile are smoother.

\section*{Acknowledgments}
This work is supported by the scientific and technological research council of Turkey under grant (TBAG:109T083) and Istanbul
University IU-BAP:6970.
\bibliographystyle{elsarticle-num}

\newpage
\begin{center}
\textbf{FIGURE CAPTIONS}
\end{center}
\vspace{1.0cm}

\noindent \textbf{Figure 1:} (Color online) Spatial distribution
of electrons at zero magnetic field and zero temperature. (a)
gated sample where potential is fixed to $V_g=-2.0V$.
 (b) and (c) etched structures, whose etching depths are 160 nm and
 240 nm from the surface of the sample, respectively. (d) etched
 and also gated (trench gated) sample: etching depth is 240 nm,
  gate voltage is $-2.0V$.\\

\noindent \textbf{Figure 2:} a) and (b) show the spatial
distribution of ISs for gated samples at 2.6 T magnetic field.
$V_g=-2.0V$ in (a) and $V_g=-5.0V$ in (b). (c) and (d) show the
spatial distribution of ISs for 160nm etched and also -2.0V
potential is applied (trench gated) sample. The magnetic field is
2.6 T in (c)
and 5.1 T in (d).\\

 \noindent \textbf{Figure 3:} Spatial distribution of the
incompressible strips for $\nu=1/3$ state at different gate
voltages. In (a) $V_g=-3.0V$, (b) $V_g=-4.0V$, (c) 160 nm etched
sample is considered at $14.4$ T (d) 160 nm etched and also
$V_g=-4.0V$ gated sample is considered at $14.4$ T.\\

\noindent \textbf{Figure 4:} Spatial distribution of the
incompressible strips for $\nu=1/3$ state in 240 nm etched sample
with different magnetic field
 values: (a) B=12.0 T, (b) B=13.2 T (c) B=14.4 T.

\end{document}